\documentclass[]{revtex4}
\usepackage{amsmath,bm,epsfig}
\pdfoutput=1
\usepackage{epstopdf}
\usepackage{subfig}
\usepackage{color}
\begin{document}
\title{Long-time evolution of sequestered CO$_2$ in porous media}
\author{Yossi Cohen$^{1}$, Daniel H. Rothman$^{1}$}
\address{$^{1}$Lorenz Center, Department of Earth Atmospheric and Planetary Sciences,
Massachusetts Institute of Technology, Cambridge, MA. 02139, USA},



\begin{abstract}
CO$_2$ sequestration in subsurface reservoirs is important for limiting atmospheric CO$_2$ concentrations. However, a complete physical picture able to predict the structure developing within the porous medium is lacking. We investigate theoretically reactive transport in the long-time evolution of carbon in the brine-rock environment.
As CO$_2$ is injected into a brine-rock environment, a carbonate-rich region is created amid brine.
Within the carbonate-rich region minerals dissolve and migrate from regions of high concentration to low concentration, along with other dissolved carbonate species.
This causes mineral precipitation at the interface between the two regions. We argue that precipitation in a small layer reduces diffusivity, and eventually causes mechanical trapping of the CO$_2$.
Consequently, only a small fraction of the CO$_2$ is converted to solid mineral; the remainder either dissolves in water or is trapped in its original form.
We also study the case of a pure CO$_2$ bubble surrounded by brine and suggest a mechanism that may lead to a carbonate-encrusted bubble due to structural diffusion.

\end{abstract}

\maketitle

\section{Introduction}
The sequestration of CO$_2$ in geological formations is widely considered to be an important approach for mitigating the rise of atmospheric CO$_2$ levels \cite{IPCC,07S,94BGP,03L,13SZ}. Deep saline aquifers and gas fields are primarily chosen for storage \cite{IPCC,03L,03BA}.
Supercritical CO$_2$ is injected into these porous media while displacing another fluid, brine \cite{13SZ,07CBT}. The CO$_2$ then gradually dissolves; however, its long-term fate remains poorly understood \cite{04BW}. Here we address the long time scale and study theoretically the evolution of the CO$_2$ after injection into the brine-rock system.

As the CO$_2$ is injected into the brine-rock environment, it initially becomes trapped, either by a physical mechanism in the presence of low permeability rocks, or by retention as a separate phase in the pore space due to interfacial tension \cite{04GBB}.
The disordered structure of the void spaces forces the injected fluid along certain paths that create regions, or bubbles, of the injected fluid amid regions of the defending fluid, and vice versa \cite{86DW,SA01,83WW,F88}. This process is known as invasion percolation when, as for the supercritical CO$_2$, the invading fluid is non-wetting. For two immiscible fluids, the fluid configuration is often determined by the structure of the rock and by surface tension effects. Further, the two-phase system can become unstable when reactant particles migrate from one phase to the other and change the chemical composition of each phase \cite{12RR,12NSYA}.
Within the high CO$_2$ phase, minerals dissolve; diffusion causes minerals and carbonate species to migrate from high concentration to low concentration regions, and a fraction of them precipitates. In nature, this process can be seen in hot springs, when a bubble of oxygen emerges from photosynthetic cyanobacteria \cite{91CRU, 00FFMP}. The high gradient of CO$_2$ between the bubble and the surrounding water leads to loss of CO$_2$ in the vicinity of the bubble, drives up the saturation level, and eventually a crust is created around the bubble. In general, mineral precipitation on a small boundary layer at the interface may lead to lower diffusivity and slower kinetics. In the carbon sequestration process, this may cause a mechanical trapping of the CO$_2$ bubble and lower the solidification rate of the carbon minerals.

Here we develop theoretical understanding of this process of mechanical phase separation. We consider two scales: At the microscale, a single CO$_2$ bubble is surrounded by brine in the void space of a porous medium. In this case, the reactions occur at the interface between the bubble and the brine, as the CO$_2$ dissolves and reduces the pH in its vicinity. The macroscale averages over many such bubbles. In this case, a high concentration of the invaded CO$_2$ changes the properties of a macroscopic region. The region becomes more acidic and no precipitation occurs unless the carbonate species migrate to a different region.

The paper begins by addressing the macroscale problem and the mathematical background of the reactive diffusion equation. We then study the mobility change in the fluid-rock system due to the precipitate minerals. Finally, we consider the microscale case of a single CO$_2$ bubble amid brine and suggest a second mechanism that may also lead to separation and self-sealing.

\section{Macroscopic Reactive Transport}
\label{rtm}
\begin{figure}
\centering
  \includegraphics[scale=1.5]{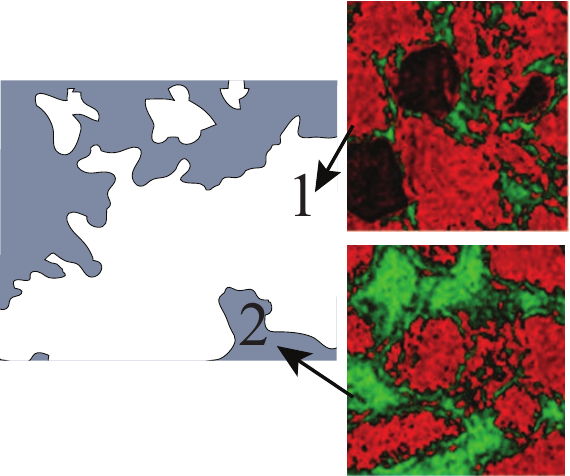}
  \caption{\textbf{Macroscale}: \emph{(Left)} Irregularly shaped regions of brine (grey) amid carbonate rich regions (white) \cite{12RR}. \textbf{Microscale:} CT image of material distribution in Frio sandstone, adapted from \cite{11STBP}. \emph{(Upper right)} Region \textbf{1} consists of CO$_2$ bubbles (black) surrounded by brine (green) and grains of the rock (red). \emph{(Lower right)} Region \textbf{2} is the brine-rock system. }\label{phases}
\end{figure}

The injection of supercritical CO$_2$ (scCO$_2$) into a porous medium, initially occupied with another fluid, i.e., brine (salty water), generates two different regions in the system \cite{12RR,13TRS}. The first region is where the CO$_2$ displaced most of the existing brine.  In this region, the void space in the porous medium is filled with bubbles of CO$_2$ and water saturated with carbonate species. The CO$_2$ dissolves rapidly in the water to reach equilibrium which results in low pH, and also a high concentration of dissolved minerals \cite{10LM,06KCHGKF}.  The second region is the intact brine-rock system, which is characterized by a low concentration of CO$_2$ and higher pH. Fig. \ref{phases} depicts the complexity of the CO$_2$-brine-rock system.

The existence of two phases and concentration gradients drives the components to migrate from one phase to another. As they diffuse, they react to reach a local equilibrium. Although pure scCO$_2$ may be clogged due to mechanical or capillary trapping \cite{09MK, IPCC}, in the presence of water it can dissolve into its ionic forms, i.e. bicarbonate and carbonic acid, until it reaches a local thermodynamical equilibrium \cite{ZW01}. These carbonate species may diffuse more easily through the brine. Within the high CO$_2$ phase, minerals dissolve because of the acidic environment. They then migrate from high concentration to low concentration regions and a fraction of them precipitates. The evolution of each component can be described by the reactive diffusion equation
\begin{equation}
\frac{\partial C_i}{\partial t} = \nabla D_i\cdot(\nabla C_i) + R_i(C_j, \mathbf{r}, t).
\label{reacdif}
\end{equation}
Here, the $C_i$, $i=1,..,5$, are the concentration of the CO$_2$, HCO$^-_3$, CO$^{2-}_3$, H$^+$ and the mineral Ca$^{2+}$, respectively.
$D_i$ is the isotropic diffusion coefficient for the $i$th component, and $R_i$ is the reaction rate defined by the carbonate system \cite{ZW01,05DM} and the dissolution and precipitation of calcite mineral. The latter can be expressed as \cite{78PWP,89CGW,11D},
\begin{equation}
\frac{\mathrm{d} M}{\mathrm{d}t}=-k_m(1- \Omega).
\label{calcite}
\end{equation}
where $M$ represents the density of precipitated calcite. The rate coefficient is defined by \cite{89CGW}
\begin{equation}
k_m=A(\mathbf{r},t)\left(k_{+1}[\mathrm{H}^+] + k_{+2}[\mathrm{CO}_2]+ k_{+3}\right),
\end{equation}
where $A(\mathbf{r},t)$ is the reactive surface area, and k$_{+i}$ are the rate constants for the forward reactions.
The saturation ratio \cite{78PWP}
\begin{equation}
\Omega=\{\mathrm{Ca}^{2+}\}\{\mathrm{CO}_3^{2-}\}/K_{sp}
\end{equation}
is defined by the ion activity product divided by the solubility constant of calcite. Details of the reactions and the values of the reaction constants can be found in ref. \cite{89CGW,08LSY}. Whether the calcite will precipitate or dissolve is determined by the value of $\Omega$; when $\Omega<1$ dissolution occurs; when $\Omega>1$ mineral precipitates, and when $\Omega=0$ the system is at equilibrium.

\section{Numerical Simulation}
In our model, we assume that there is no diffusion of the solid mineral as it nucleates and precipitates at the surface of the rock. Also, the capillary forces between the scCO$_2$ and the water prevent the mixing between the two phases; therefore the diffusion constant of CO$_2$ is also set to zero.
Since the dissolution of calcite is considerably fast compared to dissolution of other minerals, we assume that, after injection, calcite dissolves quickly until it reaches equilibrium. Only new calcite that has been precipitated can be dissolved again. Thus, the reactive surface area is set to zero, as long as there is no previous precipitation of calcite.

We solve the reactive diffusion equations (Eqs. \eqref{reacdif} and \eqref{calcite}) for each component using Galerkin finite elements method on quadratic triangular grid with a $4^{th}$ order Runge-Kutta integration scheme. The simulation starts when the carbonate system, the pH, and the dissolved calcium mineral are at local equilibrium in each phase. In region 1, the carbonate-rich brine-rock region (see Fig. \ref{phases}), we set the pH to $6$ and the total dissolved inorganic carbon to $10^{-2}$ mol/liter. In region 2, we set pH$=7$ and the carbonate concentration at $10^{-3}$ mol/liter. The simulation starts with the concentration of each one of the carbonate species and the dissolved calcium mineral at local equilibrium in each region.

\begin{figure}[htp]
\centering
\subfloat[]{
  \includegraphics[scale=0.3]{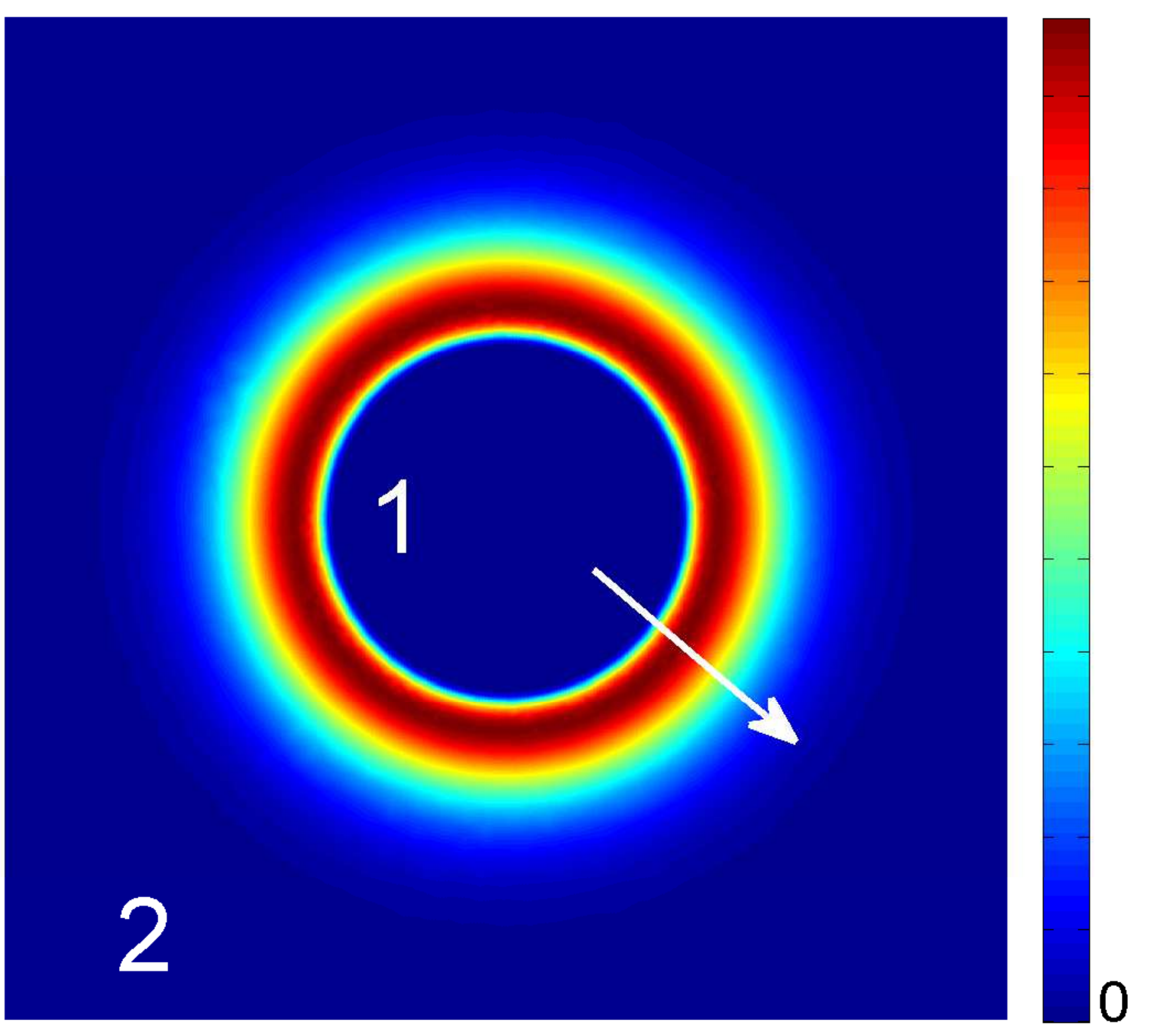}
\label{circ}
}
\hspace{0mm}
\subfloat[]{
  \includegraphics[scale=0.45]{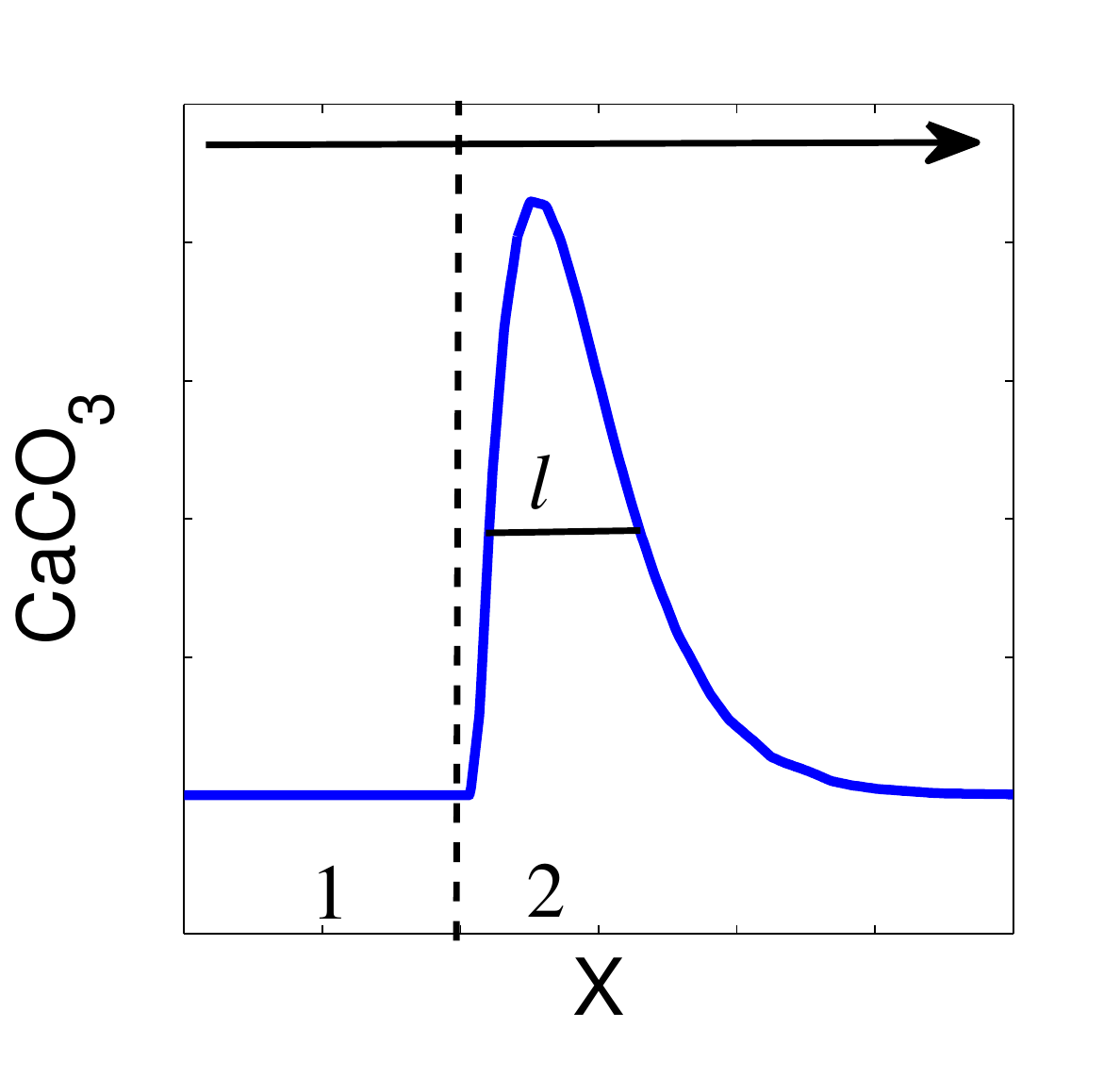}
\label{profile}
}
\caption{\textbf{a.} Mineral precipitation on the periphery of a circular domain. In the inner shape is the carbonate-rich phase (region 1) consists of high concentration of CO$_2$, amid region 2, the brine. Precipitation of minerals occurs at the interface that separates the two regions. The color indicates the accumulation of precipitated minerals in an arbitrary linear scale. \textbf{b.} The profile of the accumulated mineral (arbitrary linear scale). The dash line separates the two regions and the black arrow corresponds to the white arrow in (a).}
\end{figure}

We consider a carbonate-rich circular shape (region 1) is surrounded by brine (region 2). We find that the accumulation of calcite, shown in Fig. \ref{circ}, occurs on a small boundary layer close to the interface. The precipitation profile is approximated by the typical diffusion length $l\sim \sqrt{Dt}$, shown in Fig. \ref{profile}.

\begin{figure}[ht]
\centering
  \includegraphics[scale=0.25]{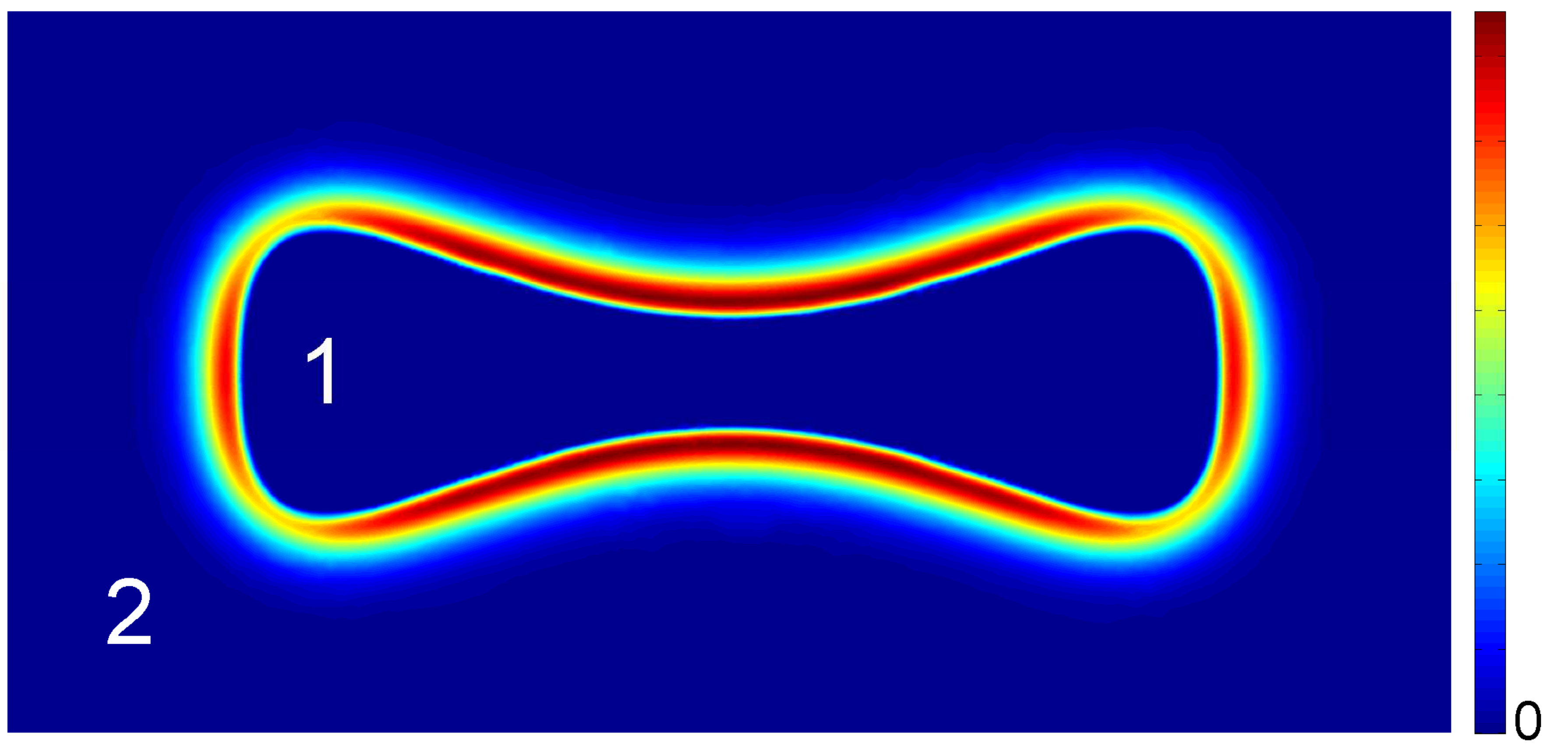}
  \caption{The same as figure \ref{circ}, but with an interface of variable curvature. Precipitation is greatest in regions of high negative curvature (concave outwards)}\label{cur}
\end{figure}

Fig. \ref{cur} shows qualitatively how the interface curvature alters the mineral precipitation. We find that the accumulated crust is highly dependent on the curvature of the interface due to the nature of a diffusion process through a curved interface \cite{03BP}; negative curvature with respect to the inner bubble of the CO$_2$ phase has maximum mineral precipitation at the interface.

Collectively, these results indicate that precipitation of mineral occurs on a small boundary layer at the interface and a carbonate crust, created in a small boundary layer, can lead to a mechanical separation of the two phases.

\section{System Size Dependence}
\label{ta}
\begin{figure}[ht]
\centering
  \includegraphics[scale=0.6]{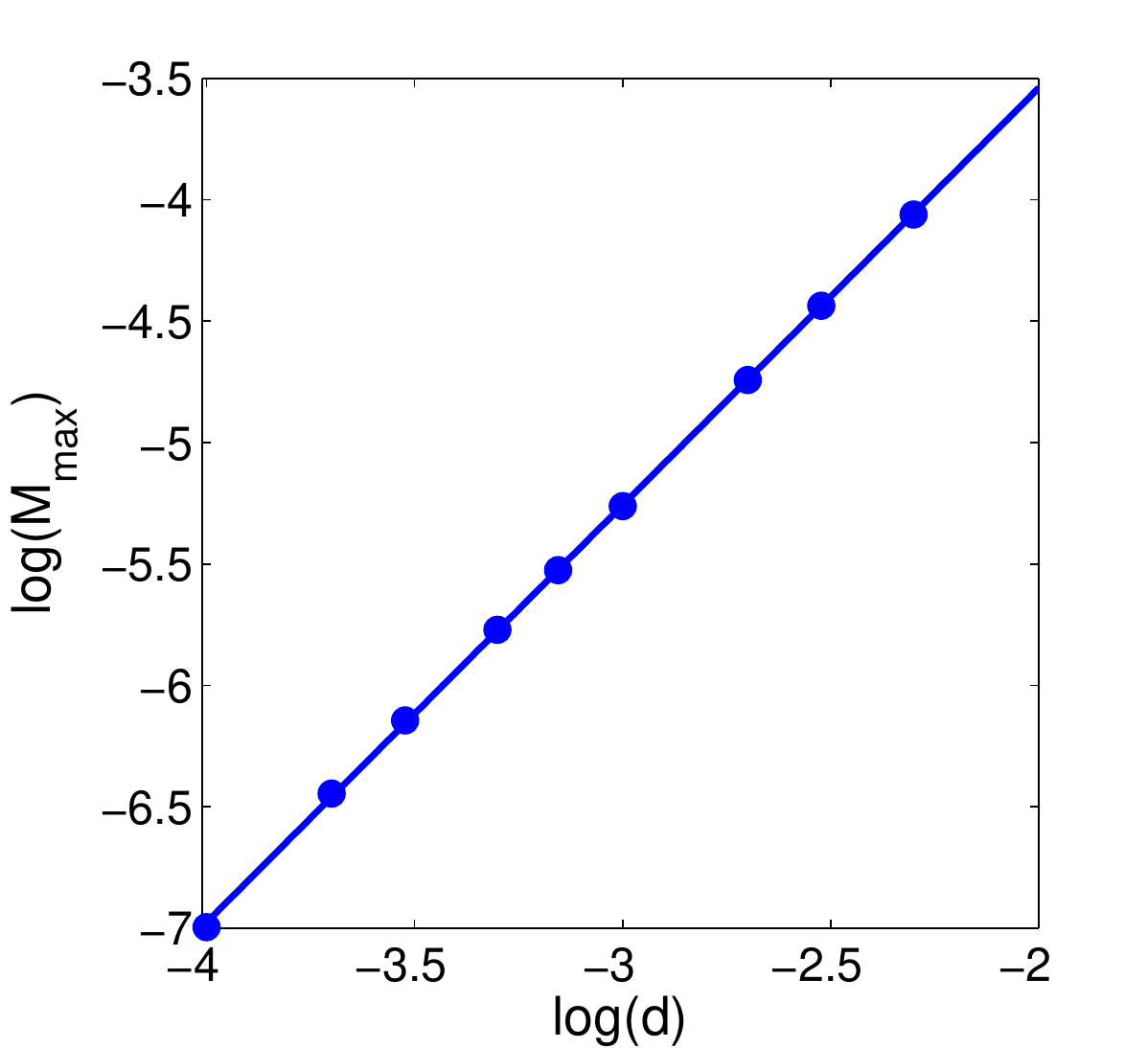}
  \caption{Maximum density of precipitated mineral with respect to system size. The slope is $n=1.72$.}\label{size}
\end{figure}

Mineral precipitation along the interface changes the porosity of the rock, decreases the effective diffusivity, and may clog existing voids \cite{09MK,IPCC}. The decrease in the mobility of the ions in the solution could bring further solidification of minerals to a halt. Whether a pathway between pores will be clogged or not is highly dependent on the local density of the accumulated mineral, which depends also on the amount of the carbonate species, and the size and the shape of the domain noted as region 1. To study how a change in the size alters the maximum mineral density, we initiate a 1-dimensional domain of size \emph{d} of region 1, and run the simulation until equilibrium. During the simulation, the mineral density reaches a maximum and then decreases as the solution at the interface become more acidic. We calculate the maximum calcite concentration $M$ as a function of $d$. We obtain a power law relationship $M\propto d^n$, where $n\simeq1.72$, as shown in Fig. \ref{size}.

The exponent $n$ can be approximated by a simple scaling analysis. Consider the reactive diffusion equation, Eq. \eqref{reacdif}. Changing the equation into dimensionless variables, $x\rightarrow \xi d$, $t\rightarrow \frac{d^2}{D} \tau$, leads to
\begin{equation}
\frac{\partial C_i}{\partial \tau} =  \frac{\partial^2 C_i}{\partial \xi^2}  +\frac{d^2}{D} R_i(C_j).
\label{reacdiffS}
\end{equation}
and from Eq. \ref{calcite}, the density of the precipitated mineral becomes
\begin{equation}
\frac{\mathrm{d}M}{\mathrm{d} \tau}=\frac{k_m d^2}{D}\left(1-\frac{A(\tau)B(\tau)}{K_{sp}}\right)
\label{calciteS}
\end{equation}
where $A(\tau)$ and $B(\tau)$ correspond to the concentration of Ca$^{2+}$ and CO$_3^{2-}$.

If $A(\tau)$ and $B(\tau)$ are purely diffusion controlled, i.e. if the reaction term in Eq. \eqref{reacdiffS} can be neglected, the accumulated precipitation scales like $d^2$. In practice, the exponent may be lower than 2 because of the contribution of the reaction terms that change the equilibrium state in our system.

\section{Effective Diffusivity}
The diffusivity in a porous medium depends on several factors which are related to the pore geometry and the effective porosity accessible by diffusion \cite{S11,89SJ}.
For simplicity, we consider the permeability reduction to be linear with the porosity loss and mineral precipitation, and the effective diffusivity $D_e$ to be linear with the porosity $\phi$:
\begin{equation}
D_e\propto\phi_t D_0, \qquad \phi_t=\left(1-\frac{\theta}{\theta_c}\right),
\label{effD}
\end{equation}
where $D_0$ is the bulk diffusivity corresponding to the initial porosity, $\phi_t$ is the porosity available for transport, $\theta=M$ is the maximum density of the precipitated mineral and $\theta_c$ is a critical accumulated density in which $D_e$ vanishes.

\begin{figure}[htp]
\centering
  \includegraphics[scale=0.6]{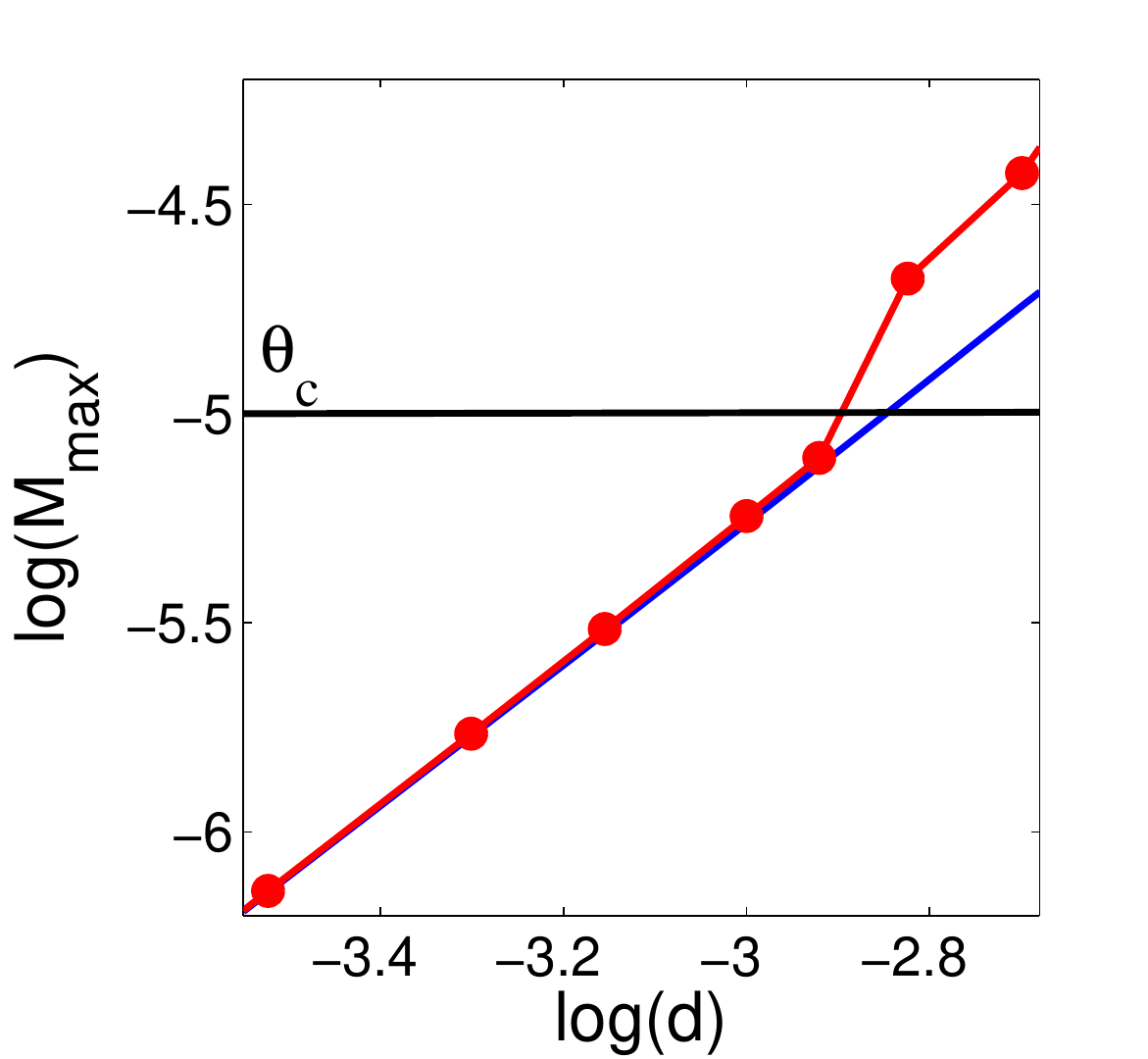}
  \caption{Maximum density of precipitated mineral \emph{M} with respect to system size \emph{d}, with clogging. \emph{Blue line}: the same as Fig. \ref{size} with $D_e=D_0$. \emph{Red squares}: The results allowing the porosity to decrease and the effective diffusivity $D_e$ to decrease as the precipitation grows. Note the enhanced precipitation when the critical porosity, $\theta_c$, is reached. At this point, particles cannot cross between the regions, and the system seeks a local chemical equilibrium.}\label{eff}
\end{figure}
As the accumulated mineral reaches the critical porosity, it changes locally the permeability and mechanical separation may occur due to diffusivity loss and self-sealing. We also observe that while the diffusion coefficient decreases, it creates even more precipitation in a smaller area close to the interface. These results, shown in Fig. \ref{eff}, are in agreement with the diffusion length scale of $l\sim\sqrt{D_et}$ from the numerical approximation and the mineral precipitation of $M\sim1/D_e$ from Eq. \eqref{calciteS}. In addition, when the system is clogged, i.e. when the effective diffusivity goes to zero, the system seeks a local chemical equilibrium.

From the numerical results, we can identify two trapping mechanisms that occur as particles migrate from one region to the other. For small carbonate-rich regions, the CO$_2$ dissolves completely into the brine and the low pH does not allow a significant precipitation of the carbonate minerals. Thus, the CO$_2$ is trapped in its dissolved forms. For larger regions, solidification of minerals stops as clogging occurs, and the developed crust separates the two regions. In both cases the total amount of the precipitated carbonate minerals is small compare to concentrations of the other carbonate species.

\section{The Microscale: a Single CO$_2$ Bubble}
\begin{figure}[htp]
\centering
\subfloat[]{
\includegraphics[scale=0.23]{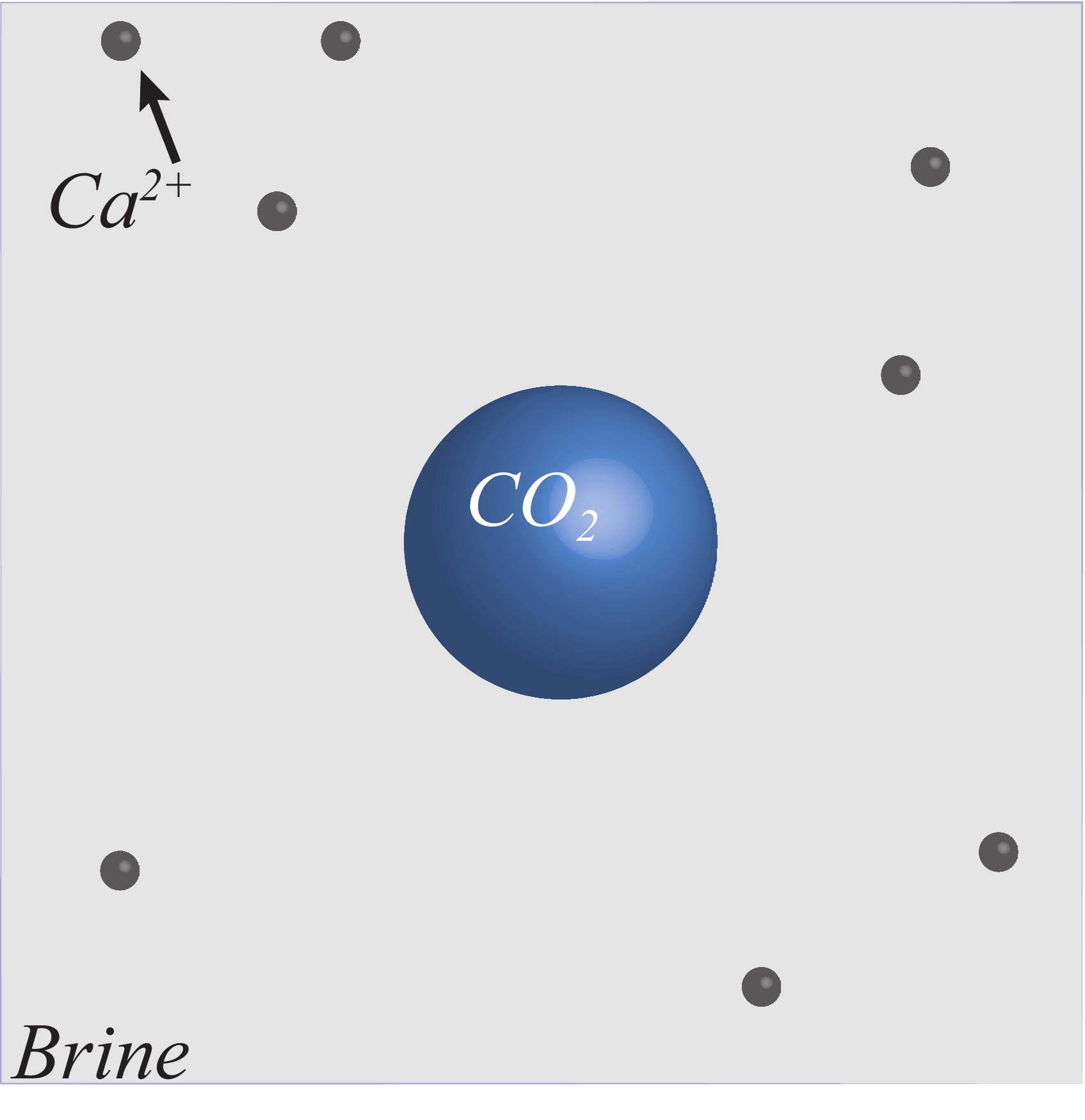}
\label{b1}
}
\subfloat[]{
\includegraphics[scale=0.23]{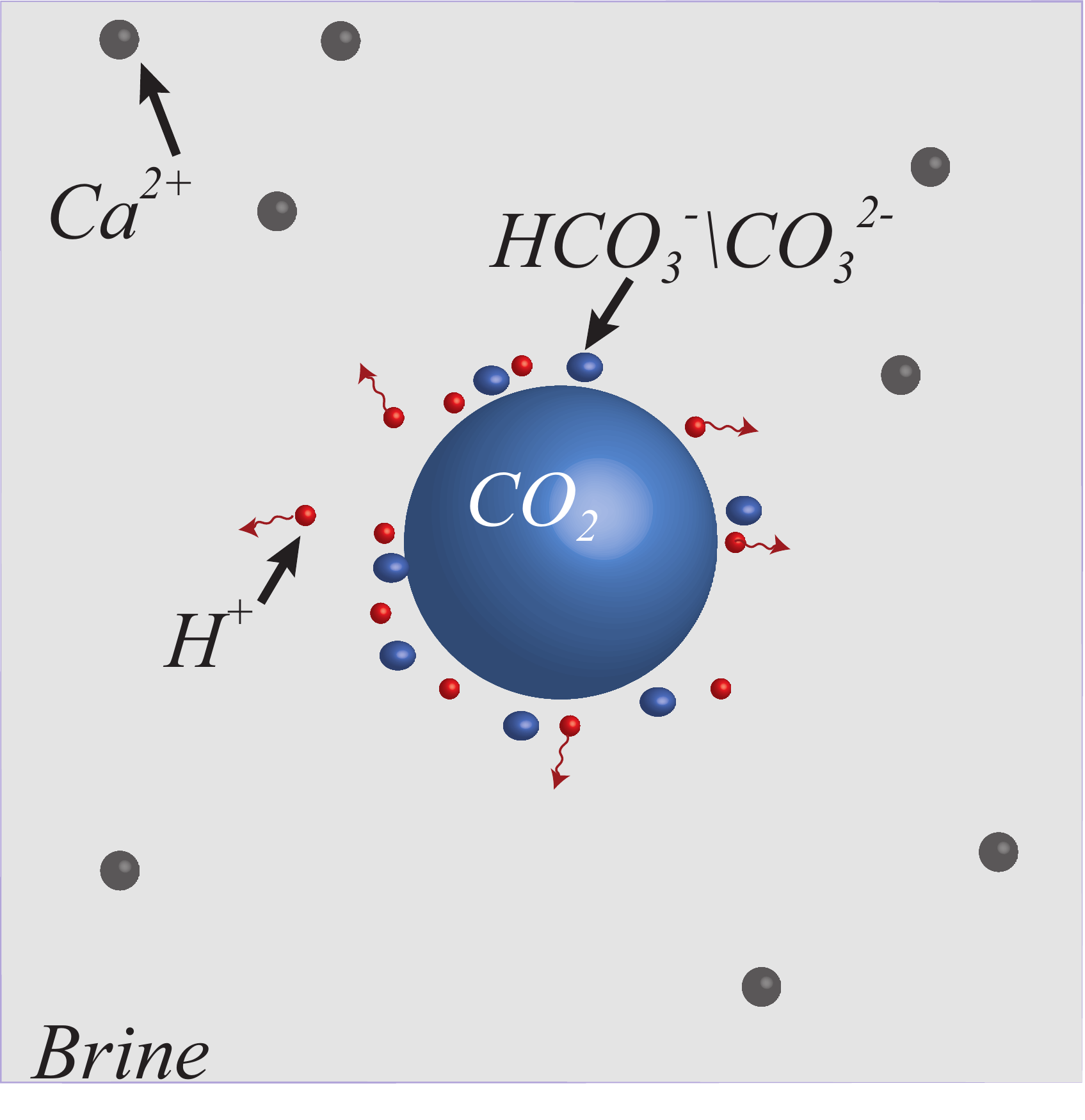}
\label{b2}
}
\hspace{0mm}
\subfloat[]{
\includegraphics[scale=0.23]{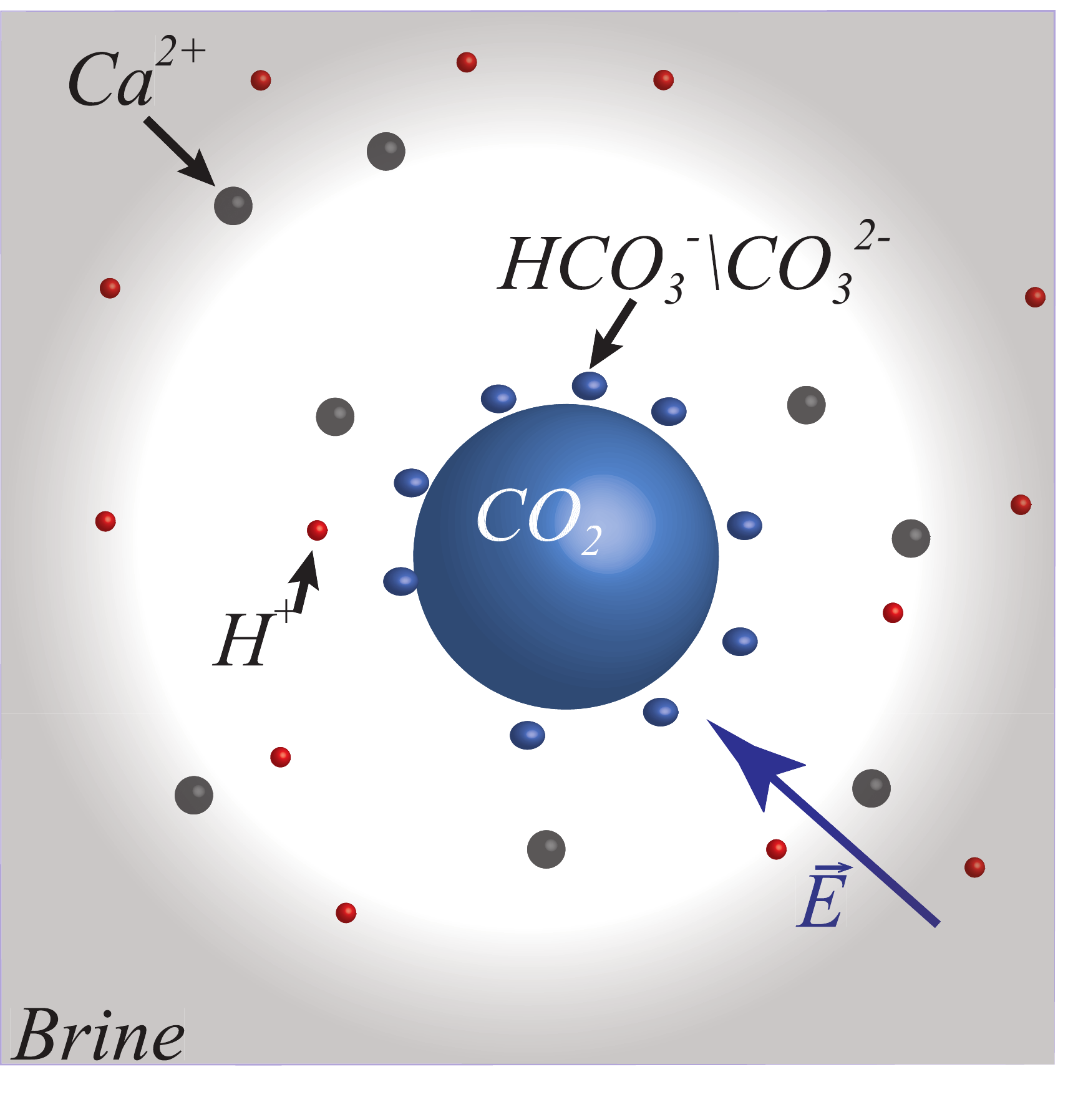}
\label{b3}
}
\subfloat[]{
\includegraphics[scale=0.23]{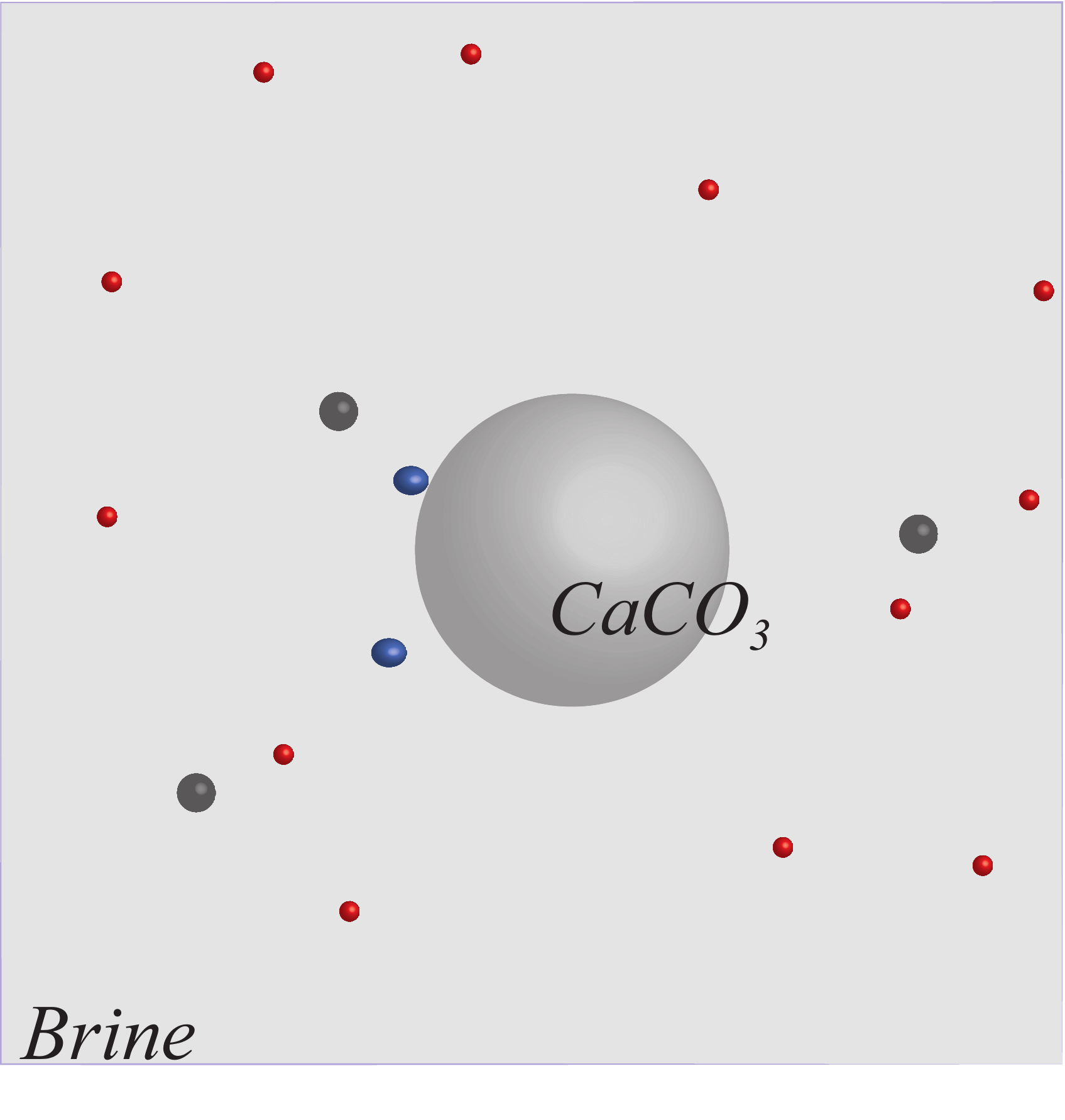}
\label{b4}
}
\caption{Evolution of a CO$_2$ bubble (not in scale): \textbf{a.} A single bubble of pure CO$_2$ surrounded by brine. \textbf{b.} At the interface, the CO$_2$ dissolves into the brine, and reduces the pH. \textbf{c.} Protons migrate faster from the interface, creating an electrostatic dipole, and attracting minerals toward the interface. \textbf{d.} Minerals precipitate due to supersaturation, and a crust is created.}\label{pco2}
\end{figure}

In this section, we discuss a mechanism that leads to a carbonate-encrusted bubble, in which a single bubble of pure CO$_2$ surrounded by brine develops a crust at the interface. Unlike the upscaled case discussed above, where each region consists of the same ingredients but with a different concentration, here we have a pure bubble (no upscaling) with only CO$_2$, either in a liquid phase or gas phase. The problem is pictured in Fig. \ref{pco2}.
In this case, no chemical reactions occur inside the CO$_2$ bubble; reactions instead occur only at the interface with the brine. The CO$_2$ then dissolves into the brine, creates charged carbonate species and reduces the pH in its vicinity. Most of the components diffuse slowly away from the interface, however, the protons in water diffuse faster due to structural diffusion, also known as the Grotthoss mechanism \cite{95A}. The concentration gradient with unequal diffusivities generates an electrical field that slows down the rapidly diffusing protons, but also attracts positively charged minerals toward the interface. (The dynamics is described by the Nernst-Planck equations \cite{97RAM,12AVMBBB}). This leads to supersaturation at the interface and mineral precipitation. This process suggest that an isolated bubble can remain stable due to a self-sealing mechanism.

\section{Discussion and Conclusion}
We have shown that in a system of reactive fluids, a gradient in the concentration between regions leads to a supersaturation at the interface and to precipitation and porosity loss. A local change in the porosity reduces permeability and may cause mechanical separation. This process is important in understanding the long-term sequestration of carbon dioxide in subsurface geological formation. We predict that a small domain of a region consisting mostly of the carbonate species in an acidic environment will dissolve completely into the brine. Larger domains are more likely to be self-sealed, and only a fraction of the carbonate species will be precipitated. For a single CO$_2$ bubble in brine solution, we suggest a mechanism that causes self-sealing due to the pH gradient at the interface of the bubble and structural diffusion.

Our results suggest that only a small fraction of the injected CO$_2$ is converted to a solid mineral. The remainder stays in its dissolved ionic form or is trapped in its original form. Whether a domain will go into dissolution trapping or mechanical separation depends on the concentration gradients, the properties of the rock and the porosity available for transport.

\section*{Acknowledgements}
We thank M. Z. Bazant, A. Kudrolli and S. R. Pride for sharing their thoughts and helpful discussions. The work was supported by the Center for Nanoscale Control of Geologic CO2, an Energy Frontier Research Center funded by the US Department of Energy, Office of Science, Office of Basic Energy Sciences under Award No. DE-AC02-05CH11231, subcontract 6896518.


\begin{thebibliography}{10}

\bibitem{IPCC}
Metz B, Davidson O, De~Coninck H, Loos M, Meyer L. 2007
\emph{{IPCC} special report on carbon dioxide capture and storage: Prepared
  by working group III of the intergovernmental panel on climate change.}
{IPCC} \textbf{4}, Cambridge University Press: Cambridge, United Kingdom and New
  York, USA..

\bibitem{07S}
Schrag DP. 2007
{Preparing to capture carbon.}
Science. \textbf{315} (5813):812--813.

\bibitem{94BGP}
Bachu S, Gunter W, Perkins E. 1994
{Aquifer disposal of {CO}$_2$: Hydrodynamic and mineral trapping.}
Energy Conversion and Management. \textbf{35}(4):269--279.

\bibitem{03L}
Lackner KS. 2003
{A guide to {CO$_2$} sequestration.}
Science. \textbf{300}(5626):1677--1678.

\bibitem{13SZ}
Song J, Zhang D. 2013
{Comprehensive Review of Caprock-Sealing Mechanisms for Geologic
  Carbon Sequestration.}
Environmental Science \& Technology. \textbf{47}(1):9--22.

\bibitem{03BA}
Bachu S, Adams J. 2003
{Sequestration of {CO}$_2$ in geological media in response to climate
  change: capacity of deep saline aquifers to sequester {CO$_2$} in solution.}
Energy Conversion and Management. \textbf{44}(20):3151--3175.

\bibitem{07CBT}
Chiquet P, Broseta D, Thibeau S. 2007
{Wettability alteration of caprock minerals by carbon dioxide.}
Geofluids. \textbf{7}(2):112--122.

\bibitem{04BW}
Baines SJ, Worden RH. 2004
{The long-term fate of {CO$_2$} in the subsurface: natural analogues
  for {CO$_2$} storage.}
Geological Society, London, Special Publications. \textbf{233}(1):59--85.

\bibitem{04GBB}
Gunter WD, Bachu S, Benson S. 2004
{The role of hydrogeological and geochemical trapping in sedimentary
  basins for secure geological storage of carbon dioxide.}
Geological Society, London, Special Publications.
  \textbf{233}(1):129--145.

\bibitem{86DW}
Dias MM, Wilkinson D. 1986
{Percolation with trapping.}
Journal of Physics A: Mathematical and General. \textbf{19}(15):3131.

\bibitem{SA01}
Stauffer D, Aharony A. 1992
\emph{Introduction to percolation theory.}
Taylor and Francis, London.

\bibitem{83WW}
Wilkinson D, Willemsen JF. 1983
{Invasion percolation: a new form of percolation theory.}
Journal of Physics A: Mathematical and General. \textbf{16}(14):3365.

\bibitem{F88}
Feder J. 1988
\emph{Fractals.}
Plenum Press, New York.

\bibitem{12RR}
Reeves D, Rothman DH. 2012
{Impact of structured heterogeneities on reactive two-phase porous
  flow.}
Physical Review E. \textbf{86}(3):031120.

\bibitem{12NSYA}
Noiriel C, Steefel CI, Yang L, Ajo-Franklin J. 2012
{Upscaling calcium carbonate precipitation rates from pore to
  continuum scale.}
Chemical Geology. \textbf{318}:60--74.

\bibitem{91CRU}
Chafetz H, Rush PF, Utech NM. 1991
{Microenvironmental controls on mineralogy and habit of {CaCO}$_3$
  precipitates: an example from an active travertine system.}
Sedimentology. \textbf{38}(1):107--126.

\bibitem{00FFMP}
Fouke BW, Farmer JD, Des~Marais DJ, Pratt L, Sturchio NC, Burns PC, et~al. 2000
{Depositional Facies and Aqueous-Solid Geochemistry of
  Travertine-Depositing Hot Springs (Angel Terrace, Mammoth Hot Springs,
  Yellowstone National Park, U.S.A.).}
Journal of Sedimentary Research. \textbf{70}(3):565--585.

\bibitem{13TRS}
Tsai PA, Riesing K, Stone HA. 2013
{Density-driven convection enhanced by an inclined boundary:
  Implications for geological {CO}$_2$ storage.}
Physical Review E. \textbf{87}(1):011003.

\bibitem{10LM}
Liu Q, Maroto-Valer MM. 2010
{Investigation of the pH effect of a typical host rock and buffer
  solution on {CO}$_2$ sequestration in synthetic brines.}
Fuel Processing Technology. \textbf{91}(10):1321--1329.

\bibitem{06KCHGKF}
Kharaka Y, Cole D, Hovorka S, Gunter W, Knauss K, Freifeld B.2006
{Gas-water-rock interactions in Frio Formation following {CO}$_2$
  injection: Implications for the storage of greenhouse gases in sedimentary
  basins.}
Geology. \textbf{34}(7):577--580.

\bibitem{11STBP}
Silin D, Tomutsa L, Benson SM, Patzek TW. 2011
{Microtomography and pore-scale modeling of two-phase fluid
  distribution.}
Transport in Porous Media. \textbf{86}(2):495--515.

\bibitem{09MK}
Matter JM, Kelemen PB. 2009
{Permanent storage of carbon dioxide in geological reservoirs by
  mineral carbonation.}
Nature Geoscience. \textbf{2}(12):837--841.

\bibitem{ZW01}
Zeebe RE, Wolf-Gladrow D. 2001
\emph{{CO}$_2$ in Seawater: Equilibrium, Kinetics, Isotopes: Equilibrium,
  Kinetics, Isotopes.}
Elsevier.

\bibitem{05DM}
Druckenmiller ML, Maroto-Valer MM. 2005
{Carbon sequestration using brine of adjusted pH to form mineral
  carbonates.}
Fuel Processing Technology. \textbf{86}(14):1599--1614.

\bibitem{78PWP}
Plummer L, Wigley T, Parkhurst D. 1978
{The kinetics of calcite dissolution in {CO}$_2$-water systems at 5
  degrees to 60 degrees C and 0.0 to 1.0 atm {CO}$_2$.}
American Journal of Science. \textbf{278}(2):179--216.

\bibitem{89CGW}
Chou L, Garrels RM, Wollast R. 1989
{Comparative study of the kinetics and mechanisms of dissolution of
  carbonate minerals.}
Chemical Geology. \textbf{78}(3):269--282.

\bibitem{11D}
DePaolo DJ. 2011
{Surface kinetic model for isotopic and trace element fractionation
  during precipitation of calcite from aqueous solutions.}
Geochimica et Cosmochimica Acta. \textbf{75}(4):1039--1056.

\bibitem{08LSY}
Li L, Steefel CI, Yang L. 2008
{Scale dependence of mineral dissolution rates within single pores and
  fractures.}
Geochimica et Cosmochimica Acta. \textbf{72}(2):360--377.

\bibitem{03BP}
Blyth M, Pozrikidis C. 2003
{Heat conduction across irregular and fractal-like surfaces.}
International journal of heat and mass transfer.
  \textbf{46}(8):1329--1339.

\bibitem{S11}
Sahimi M. 2012
\emph{Flow and transport in porous media and fractured rock: from classical
  methods to modern approaches.}
John Wiley \& Sons, Germany.

\bibitem{89SJ}
Sahimi M, Jue VL. 1989
{Diffusion of large molecules in porous media.}
Physical review letters. \textbf{62}(6):629.

\bibitem{95A}
Agmon N. 1995
{The grotthuss mechanism.}
Chemical Physics Letters. \textbf{244}(5):456--462.

\bibitem{97RAM}
Ram{\'\i}rez P, Alcaraz A, Maf{\'e} S. 1997
{Effects of pH on ion transport in weak amphoteric membranes.}
Journal of Electroanalytical Chemistry. \textbf{436}(1):119--125.

\bibitem{12AVMBBB}
Andersen MB, Van~Soestbergen M, Mani A, Bruus H, Biesheuvel P, Bazant M. 2012
{Current-induced membrane discharge.}
Physical review letters. \textbf{109}(10):108301.

\end{thebibliography}
\end{document}